\documentclass[journal]{IEEEtran}

\usepackage{amsmath}
\usepackage{amssymb}
\usepackage{amsfonts}
\usepackage{slashbox}
\usepackage{graphicx}
\usepackage{cite, amsmath, amsfonts, amssymb, psfrag, epsfig,tikz}
\usetikzlibrary{shapes,shadows,arrows}
\newtheorem{proposition}{{Proposition}}
\newtheorem{definition}{{Definition}}
\newtheorem{conjecture}{{Conjecture}}
\newtheorem{theorem}{{Theorem}}
\newtheorem{lemma}{{Lemma}}

\newtheorem{remark}{{Remark}}

\DeclareMathAlphabet{\mathpzc}{OT1}{pzc}{m}{it}
\normalsize

\ifCLASSINFOpdf
\else
\fi

\begin{document}
%
\title{Generalized Gaussian Multiterminal Source Coding in the High-Resolution Regime}
%
%
%
\author{Li Xie, Xiaolan Tu, Siyao Zhou, Jun Chen 
}

\maketitle

\begin{abstract}

A conjectural expression of the asymptotic gap between the rate-distortion function of an arbitrary generalized Gaussian multiterminal source coding system and that of its centralized counterpart in the high-resolution regime is proposed. The validity of this expression is verified when the number of sources is no more than 3.
\end{abstract}


%
\IEEEpeerreviewmaketitle

\section{Introduction}
\begin{figure}[tb]
	\centering
	\includegraphics[width=9cm]{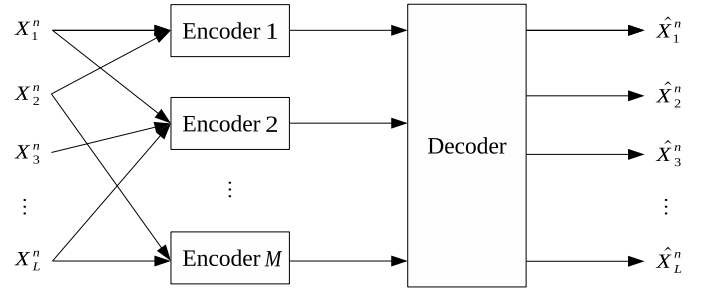}
	\caption{A generalized multiterminal source coding system with with $L$ sources and $M$ encoders. \label{fig:plot1}}
\end{figure}

%
%
%
%
\IEEEPARstart{C}{onsider} a generalized multiterminal source coding system with $L$ sources and $M$ encoders (see Fig. \ref{fig:plot1}). Each encoder compresses its observed subset of sources and forwards the compressed data to a central decoder, which attempts to reconstruct all $L$ sources based the received data to meet a prescribed distortion constraint. Such a system model arises in various scenarios. For example, the encoders and the decoder here can correspond respectively to the sensors and the fusion center in a sensor network; the flexibility of the model makes it possible to take into account the fact that the signals captured by two different sensors might share  common components. Moreover, one may interpret the encoders as a sequence of operations ordered in the temporal domain rather than some physical entities deployed in the spatial domain. For instance, the whole generalized multiterminal source coding system can be viewed as a video coding process, where at each time instant an encoding operation is performed on a batch of video frames (overlaps are allowed from batch to batch).








Two extreme cases of generalized multiterminal source coding are well known. The first one is centralized coding, where all $L$ sources are connected to a common encoder. The other one is distributed coding, where each source is connected to a different encoder. Intuitively, the optimal rate-distortion performance of any generalized multiterminal source coding system must be no superior to that of its centralized counterpart and no inferior to that of its distributed counterpart.

Special attention has been paid to the setting known as generalized (quadratic) Gaussian multiterminal source coding, where the sources are jointly Gaussian and the mean squared error distortion measure is adopted. For the centralized coding case, the rate-distortion function is given by the celebrated reverse water-filling formula \cite{CT91}.  However, for the distributed coding case, the exact characterization of the rate-distortion limit is a longstanding open problem, and so far the complete solution  has only been obtained when $L=2$ \cite{WTV08} (see also \cite{Oohama97, Oohama98, PTR04, CZBW04, Oohama05, CB08, CW08, CB082, TVW10, WCW10, YX12, YZX13, WC13, WC14, Oohama14, WXZC19} for some related results). Beyond these two extreme cases, our understanding is rather limited, and the relevant research has just started recently \cite{CEK17,CXCWW17}. Moreover, there are strong evidences that
for most generalized Gaussian multiterminal source coding systems, their rate-distortion limits might not be expressible using closed-form formulae. Indeed, the existing conclusive results for the distributed coding case are typically given in the form of semidefinite programming \cite{WCW10,YX12,YZX13,WC13,WC14,Oohama14}. Therefore, even if one manages to solve the generalized Gaussian multiterminal source coding problem completely, extracting useful insights from such a solution can still be non-trivial.

A potentially important finding of this work is that a simple picture might emerge in the high-resolution regime. Specifically, we propose a conjectural expression of the asymptotic gap between the rate-distortion function of an arbitrary generalized Gaussian multiterminal source coding system and that of its centralized counterpart. This expression delineates how the fundamental performance limit of a generalized Gaussian multiterminal source coding system depends on the source statistics and the system topology. To provide supporting evidences, we verify the validity of this expression for $L\leq 3$.




The rest of this paper is organized as follows. We state the problem definition and the main result in Section \ref{sec:conjecture}. The technical proof is presented in Section {\ref{sec:proof}}. We conclude the paper in Section \ref{sec:conclusion}.

Notation: $\mathbb{E}[\cdot]$, $\det(\cdot)$, and $\mathrm{tr}(\cdot)$ are respectively the expectation operator, the determinant operator, and the trace operator. We use $X^n$ as an abbreviation of $(X(1),\cdots,X(n))$. For any random vector $Y$ and random object $\omega$, the distortion covariance matrix incurred by the minimum mean squared error estimator of $Y$ from $\omega$ is denoted by $\mathrm{cov}(Y|\omega)$. We write $A\succ 0$ to indicate that $A$ is a positive definite matrix. Throughout this paper, little-$o$ notation $g(d)=o(f(d))$ means $\lim_{d\downarrow 0}\frac{g(d)}{f(d)}=0$, and the base of the logarithm function is $e$.













\section{Problem Definition and Main Result}\label{sec:conjecture}

Let $\{X_{\ell}(t)\}_{t=1}^\infty$, $\ell=1,\cdots,L$, be $L$ sources with $\{X_{\ell}(t)\}_{t=1}^\infty$, $\ell\in\mathcal{S}_m$, connected to encoder $m$, $m=1,\cdots,M$. We require that each source be connected to at least one encoder, and each encoder be connected to at least one source.
As a consequence, $\mathbb{S}\triangleq\{\mathcal{S}_1,\cdots,\mathcal{S}_M\}$ is a cover of $\{1,\cdots,L\}$ (in other words, $\mathbb{S}$ is a family of nonempty subsets of $\{1,\cdots,L\}$ whose union contains  $\{1,\cdots,L\}$).

In this paper, $(X_1(t),\cdots,X_L(t))$, $t=1,2,\cdots$, are assumed to be i.i.d. copies of a zero-mean Gaussian random vector $(X_1,\cdots,X_L)$ with positive definite covariance matrix $\Gamma$. The information matrix (or the precision matrix) $\Theta$ is defined as the inverse of $\Gamma$. The $(i,j)$-entries of $\Gamma$ and $\Theta$ are denoted by $\gamma_{i,j}$ and $\theta_{i,j}$, respectively,  $i,j\in\{1,\cdots,L\}$.

\begin{definition}\label{def:ratedistortion}
Given a cover $\mathbb{S}\triangleq\{\mathcal{S}_1,\cdots,\mathcal{S}_M\}$ of $\{1,\cdots,L\}$ and a positive number $d$, we say that rate $r$ is achievable if for any $\epsilon>0$, there exist encoding functions $\phi^{(n)}_{m}:\mathbb{R}^{|\mathcal{S}_m|\times n}\rightarrow\mathcal{C}^{(n)}_m$, $m=1,\cdots,M$, satisfying
\begin{align*}
&\frac{1}{n}\sum\limits_{m=1}^M\log|\mathcal{C}^{(n)}_m|\leq r+\epsilon,\\
&\frac{1}{Ln}\sum\limits_{\ell=1}^L\sum\limits_{t=1}^n\mathbb{E}[(X_{\ell}(t)-\hat{X}_{\ell}(t))^2]\leq d+\epsilon,
\end{align*}
where 
\begin{align*}
&\hat{X}^n_{\ell}\triangleq\mathbb{E}[X^n_{\ell}|\phi^{(n)}_{1}((X^n_{\ell_1})_{\ell_1\in\mathcal{S}_1}),\cdots,\phi^{(n)}_M((X^n_{\ell_M})_{\ell_M\in\mathcal{S}_M})],\\
&\hspace{2.5in} \ell=1,\cdots,L.
\end{align*}
The minimum of such $r$ is denoted by $r_{\mathbb{S}}(d)$. We shall refer to $r_{\mathbb{S}}(\cdot)$ as the rate-distortion function (or more precisely, the sum-rate-distortion function) of the generalized Gaussian multiterminal source coding system associated with $\mathbb{S}$.
\end{definition}
\begin{remark}\label{rem:remark2}
Let $\mathbb{S}$ and $\mathbb{S}'$ be two covers of $\{1,\cdots,L\}$. We say that $\mathbb{S}'$ dominates $\mathbb{S}$ if for any $\mathcal{S}\in\mathbb{S}$, there exists some $\mathcal{S}'\in\mathbb{S}'$ such that $\mathcal{S}\subseteq\mathcal{S}'$. It is clear that
\begin{align}
r_{\mathbb{S}}(d)\geq r_{\mathbb{S}'}(d),\quad d>0,\label{eq:dominance}
\end{align}
if $\mathbb{S}'$ dominates\footnote{For example, $\{\{1,2\},\{1,3\}\}$ dominates $\{\{1,2\},\{3\}\}$, but is dominated by $\{\{1,2\},\{1,3\},\{2,3\}\}$. Moreover, every cover of $\{1,\cdots,L\}$ dominates $\{\{1\},\cdots,\{L\}\}$, but is dominated by $\{\{1,\cdots,L\}\}$.} $\mathbb{S}$ because each encoder in the system associated with $\mathbb{S}$ is functionally realizable by some encoder in the system associated with $\mathbb{S}'$ that is connected to the same or more sources. Two covers $\mathbb{S}$ and $\mathbb{S}'$ are said to be equivalent\footnote{For example, $\{\{1,2\},\{2,3\}\}$ and $\{\{1\},\{1,2\},\{2,3\}\}$ are equivalent.} if  they dominate each other. For two equivalent covers $\mathbb{S}$ and $\mathbb{S}'$, we have 
\begin{align}
r_{\mathbb{S}}(d)= r_{\mathbb{S}'}(d),\quad d>0.\label{eq:equivalence}
\end{align}
A cover is said to be non-redundant\footnote{For example, $\{\{1,2\},\{2,3\}\}$ is a non-redundant cover of $\{1,2,3\}$ whereas $\{\{1\},\{1,2\},\{2,3\}\}$ is a redundant cover (since $\{1\}$ is contained in $\{1,2\}$).} if none of its elements is contained in another. 
It is easy to show that there exists a unique non-redundant cover among all equivalent ones.

\end{remark}

\begin{figure*}[tb]
	\centering
	\includegraphics[width=18cm]{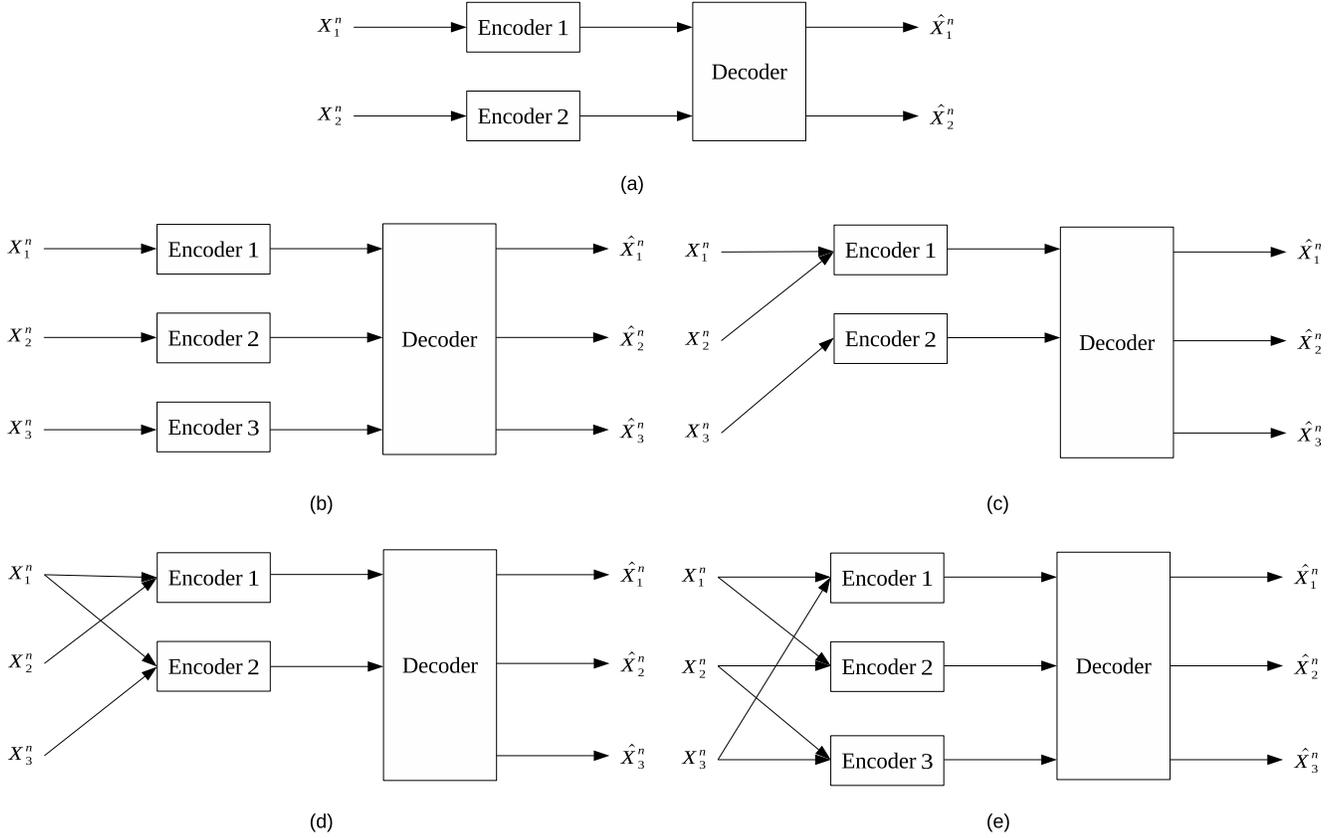}
	\caption{Examples of generalized multiterminal source coding systems: (a) $L=2$ and $\mathbb{S}=\{\{1\},\{2\}\}$, (b) $L=3$ and $\mathbb{S}=\{\{1\},\{2\},\{3\}\}$, (c) $L=3$ and $\mathbb{S}=\{\{1,2\},\{3\}\}$, (d) $L=3$ and $\mathbb{S}=\{\{1,2\},\{1,3\}\}$, (e) $L=3$ and $\mathbb{S}=\{\{1,2\},\{1,3\},\{2,3\}\}$. \label{fig:plot2}}
\end{figure*}

Let $r_C(\cdot)$ and $r_D(\cdot)$ denote the rate-distortion functions for the centralized coding case (i.e., $\mathbb{S}=\{\{1,\cdots,L\}\}$) and the distributed coding case (i.e., $\mathbb{S}=\{\{1\},\cdots,\{L\}\}$), respectively. In view of Remark \ref{rem:remark2}, we have
\begin{align}
r_C(d)\leq r_{\mathbb{S}}(d)\leq r_D(d),\quad d>0,\label{eq:inbetween}
\end{align}
for any cover $\mathbb{S}$ of $\{1,\cdots,L\}$. A result by Zamir and Berger \cite{ZB99} (see also \cite{WXZWC19} for a related result) indicates that
\begin{align*}
\lim\limits_{d\downarrow0}r_{D}(d)-r_C(d)=0,
\end{align*}
which, together with (\ref{eq:inbetween}), implies
\begin{align}
\lim\limits_{d\downarrow0}r_{\mathbb{S}}(d)-r_C(d)=0\label{eq:ZB}
\end{align}
for any cover $\mathbb{S}$ of $\{1,\cdots,L\}$. However, (\ref{eq:ZB}) falls short of capturing the dependency of $r_{\mathbb{S}}(d)$ on $\Gamma$ (or equivalently, $\Theta$) and $\mathbb{S}$. The following conjecture aims to provide a characterization of the asymptotic gap between $r_{\mathbb{S}}(d)$ and $r_C(d)$ in the high-resolution regime that is more informative than (\ref{eq:ZB}).



\begin{conjecture}\label{conj:conjecture1}
For any cover $\mathbb{S}$ of $\{1,\cdots,L\}$,
\begin{align}
r_{\mathbb{S}}(d)-r_C(d)=\frac{1}{2}\sum\limits_{(i,j)\in\mathcal{E}(\mathbb{S})}\theta^2_{i,j}d^2+o(d^2),\label{eq:gap}
\end{align}
where $\mathcal{E}(\mathbb{S})\triangleq\{(i,j): 1\leq i<j\leq L\mbox{ and }\{i,j\}\nsubseteq\mathcal{S}\mbox{ for all }\mathcal{S}\in\mathbb{S}\}$.
\end{conjecture}

\begin{remark}
	It is easy to verify that
	\begin{align*}
	\sum\limits_{(i,j)\in\mathcal{E}(\mathbb{S})}\theta^2_{i,j}\geq \sum\limits_{(i,j)\in\mathcal{E}(\mathbb{S}')}\theta^2_{i,j}
	\end{align*}
	if $\mathbb{S}'$ dominates $\mathbb{S}$,  and
	\begin{align*}
	\sum\limits_{(i,j)\in\mathcal{E}(\mathbb{S})}\theta^2_{i,j}= \sum\limits_{(i,j)\in\mathcal{E}(\mathbb{S}')}\theta^2_{i,j}
	\end{align*}
	if $\mathbb{S}$ and $\mathbb{S}'$ are equivalent. Therefore, Conjecture \ref{conj:conjecture1} is consistent with (\ref{eq:dominance}) and (\ref{eq:equivalence}).
\end{remark}

\begin{remark}
Note that $r_C(\cdot)$ is given by the reverse water-filling formula \cite{CT91}. Specifically, we have
\begin{align}
r_C(d)=\frac{1}{2}\sum\limits_{\ell=1}^L\log\Big(\frac{\lambda_{\ell}}{\min\{\delta,\lambda_{\ell}\}}\Big),\quad d>0,\label{eq:waterfilling}
\end{align}
where $\lambda_1,\cdots,\lambda_L$ are the eigenvalues of $\Gamma$, and $\delta$ is the unique solution to
\begin{align*}
\sum\limits_{\ell=1}^L\min\{\delta,\lambda_{\ell}\}=\min\{Ld,\mathrm{tr}(\Gamma)\}.
\end{align*}
In light of (\ref{eq:waterfilling}) and the fact that $\det(\Gamma)=\prod_{\ell=1}^L\lambda_{\ell}$,
\begin{align}
r_C(d)=\frac{1}{2}\log\Big(\frac{\det(\Gamma)}{d^L}\Big),\quad d\in(0,\min\{\lambda_1,\cdots,\lambda_L\}).\label{eq:shannonlowerbound}
\end{align}
As a consequence, (\ref{eq:gap}) can be written alternatively as
\begin{align*}
r_{\mathbb{S}}(d)=\frac{1}{2}\log\Big(\frac{\det(\Gamma)}{d^L}\Big)+\frac{1}{2}\sum\limits_{(i,j)\in\mathcal{E}(\mathbb{S})}\theta^2_{i,j}d^2+o(d^2),
\end{align*}
which provides conjecturally an explicit asymptotic expression of $r_{\mathbb{S}}(d)$ in the high-resolution regime. 
\end{remark}

The main contribution of this work is the following result.
\begin{theorem}\label{thm:theorem1}
Conjecture \ref{conj:conjecture1} is true for $L\leq 3$.
\end{theorem}
\begin{IEEEproof}
In view of Remark \ref{rem:remark2} and the fact that Conjecture \ref{conj:conjecture1} is trivially true if  $\mathbb{S}=\{\{1,\cdots,L\}\}$, it suffices to consider (possibly through relabelling\footnote{For example, if we relabel 1 as 2, 2 as 3, and 3 as 1, then $\{\{1,3\},\{2,3\}\}$ becomes $\{\{1,2\},\{1,3\}\}$. Clearly, it suffices to consider one of them for the purpose of proving Theorem \ref{thm:theorem1}.}) the following cases:
\begin{enumerate}
\item $L=2$ and $\mathbb{S}=\{\{1\},\{2\}\}$ (see Fig. \ref{fig:plot2} (a)),

\item $L=3$ and $\mathbb{S}=\{\{1\},\{2\},\{3\}\}$ (see Fig. \ref{fig:plot2} (b)),

\item $L=3$ and $\mathbb{S}=\{\{1,2\},\{3\}\}$ (see Fig. \ref{fig:plot2} (c)),

\item $L=3$ and $\mathbb{S}=\{\{1,2\},\{1,3\}\}$ (see Fig. \ref{fig:plot2} (d)),

\item $L=3$ and $\mathbb{S}=\{\{1,2\},\{1,3\},\{2,3\}\}$ (see Fig. \ref{fig:plot2} (e)).
\end{enumerate}
The details can be found in Section \ref{sec:proof}.
\end{IEEEproof}

\section{Proof of Theorem \ref{thm:theorem1}}\label{sec:proof}
\subsection{$L=2$ and $\mathbb{S}=\{\{1\},\{2\}\}$}

\begin{lemma}\label{lem:1}
For $d$ sufficiently close to 0, 
\begin{align*}
r_{\{\{1\},\{2\}\}}(d)&=\frac{1}{2}\log\Big(\frac{\det(\Gamma)}{2d^2}\Big(1+\sqrt{1+4\theta^2_{1,2}d^2}\Big)\Big).
\end{align*}
\end{lemma}
\begin{IEEEproof}
See Appendix \ref{app:1}.
\end{IEEEproof}

In view of (\ref{eq:shannonlowerbound}) and Lemma \ref{lem:1}, 
\begin{align*}
r_{\{\{1\},\{2\}\}}(d)-r_C(d)=\frac{1}{2}\log\Big(\frac{1}{2}+\frac{1}{2}\sqrt{1+4\theta^2_{1,2}d^2}\Big)
\end{align*}
for $d$ sufficiently close to 0. It can be verified that
\begin{align*}
&\frac{1}{2}\log\Big(\frac{1}{2}+\frac{1}{2}\sqrt{1+4\theta^2_{1,2}d^2}\Big)\\
&=\frac{1}{2}\log\Big(1+\theta^2_{1,2}d^2+o(d^2)\Big)\\
&=\frac{1}{2}\theta^2_{1,2}d^2+o(d^2),
\end{align*}
which is the desired result.

\subsection{$L=3$ and $\mathbb{S}=\{\{1\},\{2\},\{3\}\}$}

\begin{lemma}\label{lem:2}
For $d$ sufficiently close to 0,
\begin{align}
&r_{\{\{1\},\{2\},\{3\}\}}(d)=\min\limits_{\Xi}\frac{1}{2}\log\Big(\frac{\det(\Gamma)}{\det(D)}\Big)\label{eq:minimizer1}\\
&\hspace{0.55in}\mbox{subject to}\quad \Xi\succ 0,\nonumber\\
&\hspace{1.305in} \xi_{i,j}=0,\quad i\neq j,\nonumber\\
&\hspace{1.305in} \mathrm{tr}(D)\leq 3d,\nonumber
\end{align}
where $\xi_{i,j}$ denotes the $(i,j)$-entry of $\Xi$, $i,j\in\{1,2,3\}$, and $D\triangleq(\Theta+\Xi^{-1})^{-1}$.
\end{lemma}
\begin{IEEEproof}
This result can be deduced from \cite[Theorem 5]{WCW10}.
\end{IEEEproof}

According to \cite[Theorem 8]{WC14}, for $d$ sufficiently close to 0, we can find a positive definite diagonal matrix $\Xi$ such that $d_{\ell,\ell}=d$, $\ell=1,2,3$, where $d_{i,j}$ denotes the $(i,j)$-entry of $D$, $i,j\in\{1,2,3\}$; clearly,
\begin{align}
&r_{\{\{1\},\{2\},\{3\}\}}(d)-r_C(d)\nonumber\\
&\leq\frac{1}{2}\log\Big(\frac{\det(\Gamma)}{\det(D)}\Big)-\frac{1}{2}\log\Big(\frac{\det(\Gamma)}{d^3}\Big)\nonumber\\
&=\frac{1}{2}\log\Big(\frac{d^3}{\det(D)}\Big)\label{eq:upperbound1}
\end{align}
for this specifically constructed $D$.
Since
\begin{align}
(\theta_{\ell,\ell}+\xi^{-1}_{\ell,\ell})^{-1}\leq d_{\ell,\ell}\leq(\gamma^{-1}_{\ell,\ell}+\xi^{-1}_{\ell,\ell})^{-1},\quad\ell=1,2,3,\label{eq:lowerupper}
\end{align}
it follows that $\xi_{\ell,\ell}=d+o(d)$, $\ell=1,2,3$.
When the entries of $\Xi$ are sufficiently close to 0, we have
\begin{align}
D=\Xi-\Xi\Theta\Xi+\sum\limits_{n=2}^\infty(-1)^n(\Xi\Theta)^n\Xi.\label{eq:expansion}
\end{align}
It can be verified that
\begin{align*}
d_{i,j}&=-\theta_{i,j}\xi_{i,i}\xi_{j,j}+o(d^2)\\
&=-\theta_{i,j}d^2+o(d^2),\quad i\neq j,
\end{align*}
which, together with the fact that $d_{\ell}=d$, $\ell=1,2,3$, implies
\begin{align}
\det(D)=d^3-(\theta^2_{1,2}+\theta^2_{1,3}+\theta^2_{2,3})d^5+o(d^5).\label{eq:detD1}
\end{align}
Substituting (\ref{eq:detD1}) into (\ref{eq:upperbound1}) and invoking the asymptotic formula $\log(1-x)=-x+o(x)$ gives
\begin{align}
&r_{\{\{1\},\{2\},\{3\}\}}(d)-r_C(d)\nonumber\\
&\leq\frac{1}{2}(\theta^2_{1,2}+\theta^2_{1,3}+\theta^2_{2,3})d^2+o(d^2).\label{eq:firsthalf}
\end{align}

It remains to show that the above upper bound is actually tight. Let $D^*\triangleq(\Theta+(\Xi^*)^{-1})^{-1}$, where $\Xi^*$ is the minimizer of the optimization problem in (\ref{eq:minimizer1}). Denote the $(i,j)$-entries of $\Xi^*$ and $D^*$ by $\xi^*_{i,j}$ and $d^*_{i,j}$, respectively, $i,j\in\{1,2,3\}$. Clearly, for $d$ sufficiently close to 0,
\begin{align}
&r_{\{\{1\},\{2\},\{3\}\}}(d)-r_C(d)\nonumber\\
&=\frac{1}{2}\log\Big(\frac{\det(\Gamma)}{\det(D^*)}\Big)-\frac{1}{2}\log\Big(\frac{\det(\Gamma)}{d^3}\Big)\nonumber\\
&=\frac{1}{2}\log\Big(\frac{d^3}{\det(D^*)}\Big).\label{eq:lowerbound1}
\end{align}
Since 
$d^*_{1,1}+d^*_{2,2}+d^*_{3,3}=\mathrm{tr}(D^*)\leq 3d$ and $d^*_{\ell,\ell}>0$, $\ell=1,2,3$,
it follows that $d^*_{1,1}d^*_{2,2}d^*_{3,3}\leq d^3$; moreover, we must have $d^*_{\ell,\ell}=d+o(d)$, $\ell=1,2,3$, because\footnote{By Hadamard's inequality, $\det(D^*)\leq d^*_{1,1}d^*_{2,2}d^*_{3,3}$. Under the constraints $d^*_{1,1}+d^*_{2,2}+d^*_{3,3}\leq 3d$ and $d^*_{\ell,\ell}>0$, $\ell=1,2,3$, the ratio of $d^3$ to $d^*_{1,1}d^*_{2,2}d^*_{3,3}$  converges to 1 as $d\downarrow 0$ if and only if $d^*_{\ell,\ell}=d+o(d)$, $\ell=1,2,3$.} otherwise
\begin{align*}
\limsup\limits_{d\downarrow 0} r_{\{\{1\},\{2\},\{3\}\}}(d)-r_C(d)>0,
\end{align*}
wich is contradictory to (\ref{eq:firsthalf}). It can be shown by leveraging (\ref{eq:lowerupper}) and (\ref{eq:expansion}) that $\xi^*_{\ell,\ell}=d+o(d)$, $\ell=1,2,3$, and $d^*_{i,j}=-\theta_{i,j}d^2+o(d^2)$, $i\neq j$. Now one can readily verify that
\begin{align}
\det(D^*)&=d^*_{1,1}d^*_{2,2}d^*_{3,3}-(d^*_{1,2})^2d^*_{3,3}-(d^*_{1,3})^2d^*_{2,2}\nonumber\\
&\quad-(d^*_{2,3})^2d^*_{1,1}+o(d^5)\nonumber\\
&\leq d^3-(d^*_{1,2})^2d^*_{3,3}-(d^*_{1,3})^2d^*_{2,2}-(d^*_{2,3})^2d^*_{1,1}\nonumber\\
&\quad+o(d^5)\nonumber\\
&= d^3-(\theta^2_{1,2}+\theta^2_{1,3}+\theta^2_{2,3})d^5+o(d^5).\label{eq:detD2}
\end{align}
Substituting (\ref{eq:detD2}) into (\ref{eq:lowerbound1}) and invoking the asymptotic formula $\log(1-x)=-x+o(x)$ gives
\begin{align*}
&r_{\{\{1\},\{2\},\{3\}\}}(d)-r_C(d)\nonumber\\
&\geq\frac{1}{2}(\theta^2_{1,2}+\theta^2_{1,3}+\theta^2_{2,3})d^2+o(d^2).
\end{align*}
This completes the proof of Lemma \ref{lem:2}.


\subsection{$L=3$ and $\mathbb{S}=\{\{1,2\},\{3\}\}$}

\begin{lemma}\label{lem:3}
For $d$ sufficiently close to 0,
\begin{align}
&\hspace{0.1in}r_{\{\{1, 2\},\{3\}\}}(d)=\min\limits_{\Xi}\frac{1}{2}\log\Big(\frac{\det(\Gamma)}{\det(D)}\Big)\label{eq:minimizer2}\\
&\hspace{0.55in}\mbox{subject to}\quad \Xi\succ 0,\nonumber\\
&\hspace{1.305in}\xi_{1,3}=\xi_{2,3}=0,\nonumber\\
&\hspace{1.305in} \mathrm{tr}(D)\leq 3d,\nonumber
\end{align}
where $D$ is defined as in Lemma \ref{lem:2}, and $\Xi$ is a symmetric matrix with its $(i,j)$-entry denoted by $\xi_{i,j}$, $i,j\in\{1,2,3\}$.
\end{lemma}
\begin{IEEEproof}
This result can be deduced from \cite[Theorem 9]{WC13} and \cite[Theorem 9]{WC14}.
\end{IEEEproof}

According to \cite[Theorem 8]{WC14}, for $d$ sufficiently close to 0, we can find a positive definite symmetric matrix $\Xi$ with $\xi_{1,3}=\xi_{2,3}=0$ such that $d_{\ell,\ell}=d$, $\ell=1,2,3$, and $d_{1,2}=0$, where $d_{i,j}$ denotes the $(i,j)$-entry of $D$, $i,j\in\{1,2,3\}$; clearly,
\begin{align}
&r_{\{\{1,2\},\{3\}\}}(d)-r_C(d)\nonumber\\
&\leq\frac{1}{2}\log\Big(\frac{\det(\Gamma)}{\det(D)}\Big)-\frac{1}{2}\log\Big(\frac{\det(\Gamma)}{d^3}\Big)\nonumber\\
&=\frac{1}{2}\log\Big(\frac{d^3}{\det(D)}\Big)\label{eq:upperbound2}
\end{align}
for this specifically constructed $D$. In view of (\ref{eq:lowerupper}),  we must have $\xi_{\ell,\ell}=d+o(d)$, $\ell=1,2,3$, which, together with the fact that $\Xi$ is a positive definite symmetric matrix, implies $\xi_{1,2}\leq d+o(d)$. It can be verified by leveraging (\ref{eq:expansion}) that
\begin{align}
d_{1,2}&=\xi_{1,2}-\theta_{1,1}\xi_{1,1}\xi_{1,2}-\theta_{1,2}\xi^2_{1,2}-\theta_{1,2}\xi_{1,1}\xi_{2,2}\nonumber\\
&\quad-\theta_{2,2}\xi_{1,2}\xi_{2,2}+o(d^2),\label{eq:d12}\\
d_{1,3}&=-\theta_{1,3}\xi_{1,1}\xi_{3,3}-\theta_{2,3}\xi_{1,2}\xi_{3,3}+o(d^2),\nonumber\\
d_{2,3}&=-\theta_{1,3}\xi_{1,2}\xi_{3,3}-\theta_{2,3}\xi_{2,2}\xi_{3,3}+o(d^2).\nonumber
\end{align}
Since $d_{1,2}=0$ and $\xi_{\ell,\ell}=d+o(d)$, $\ell=1,2,3$, it follows by (\ref{eq:d12}) that $\xi_{1,2}=\theta_{1,2}d^2+o(d^2)$. Now one can easily show 
\begin{align*}
&d_{1,3}=-\theta_{1,3}d^2+o(d^2),\\
&d_{2,3}=-\theta_{2,3}d^2+o(d^2),
\end{align*}
which, in conjunction with the fact that $d_{\ell,\ell}=d$, $\ell=1,2,3$, and $d_{1,2}=0$, implies
\begin{align}
\det(D)=d^3-(\theta^2_{1,3}+\theta^2_{2,3})d^5+o(d^5).\label{eq:detD3}
\end{align}
Substituting (\ref{eq:detD3}) into (\ref{eq:upperbound2}) and invoking the asymptotic formula $\log(1-x)=-x+o(x)$ gives
\begin{align}
r_{\{\{1,2\},\{3\}\}}(d)-r_C(d)\leq\frac{1}{2}(\theta^2_{1,3}+\theta^2_{2,3})d^2+o(d^2).\label{eq:secondhalf}
\end{align}

It remains to show that the above upper bound is actually tight. Let $D^*\triangleq(\Theta+(\Xi^*)^{-1})^{-1}$, where $\Xi^*$ is the minimizer of the optimization problem in (\ref{eq:minimizer2}).  Clearly, for $d$ sufficiently close to 0,
\begin{align}
&r_{\{\{1,2\},\{3\}\}}(d)-r_C(d)\nonumber\\
&=\frac{1}{2}\log\Big(\frac{\det(\Gamma)}{\det(D^*)}\Big)-\frac{1}{2}\log\Big(\frac{\det(\Gamma)}{d^3}\Big)\nonumber\\
&=\frac{1}{2}\log\Big(\frac{d^3}{\det(D^*)}\Big).\label{eq:lowerbound2}
\end{align}
Denote the $(i,j)$-entries of $\Xi^*$ and $D^*$ by $\xi^*_{i,j}$ and $d^*_{i,j}$, respectively, $i,j\in\{1,2,3\}$. It is easy to see that
 $d^*_{1,1}d^*_{2,2}d^*_{3,3}\leq d^3$,    $d^*_{\ell,\ell}=d+o(d)$, $\ell=1,2,3$, $\xi^*_{\ell,\ell}=d+o(d)$, $\ell=1,2,3$, and $\xi^*_{1,2}\leq d+o(d)$. Moreover, we must have
 \begin{align*}
 \limsup\limits_{d\downarrow 0}\frac{|\xi^*_{1,2}|}{d^2}<\infty
 \end{align*}
since otherwise
 \begin{align*}
 &\limsup\limits_{d\downarrow 0}\frac{r_{\{\{1,2\},\{3\}\}}(d)-r_C(d)}{d^2}\\
 &=\limsup\limits_{d\downarrow 0}\frac{1}{2d^2}\log\Big(\frac{d^3}{\det(D^*)}\Big)\\
 &=\limsup\limits_{d\downarrow 0}\frac{1}{2d^2}\log\Big(\frac{d^3}{d^*_{1,1}d^*_{2,2}d^*_{3,3}-(d^*_{1,2})^2d^*_{3,3}}\Big)\\
 &\geq\limsup\limits_{d\downarrow 0}\frac{1}{2d^2}\log\Big(\frac{d^3}{d^3-(d^*_{1,2})^2d^*_{3,3}}\Big)\\
 &=\limsup\limits_{d\downarrow 0}\frac{(d^*_{1,2})^2d^*_{3,3}}{2d^5}\\
 &=\limsup\limits_{d\downarrow 0}\frac{(\xi^*_{1,2})^2}{2d^4}\\
 &=\infty,
 \end{align*}
 wich is contradictory to (\ref{eq:secondhalf}). This along with the fact that $\xi^*_{\ell}=d+o(d)$, $\ell=1,2,3$, implies 
 \begin{align*}
 &d^*_{1,3}=-\theta_{1,3}d^2+o(d^2),\\
 &d^*_{2,3}=-\theta_{2,3}d^2+o(d^2).
 \end{align*}
 Now it can be verified that
 \begin{align}
 \det(D^*)&=d^*_{1,1}d^*_{2,2}d^*_{3,3}-(d^*_{1,2})^2d^*_{3,3}-(d^*_{1,3})^2d^*_{2,2}\nonumber\\
 &\quad-(d^*_{2,3})^2d^*_{1,1}+o(d^5)\nonumber\\
 &\leq d^3-(d^*_{1,3})^2d^*_{2,2}-(d^*_{2,3})^2d^*_{1,1}+o(d^5)\nonumber\\
 &=d^3-(\theta^2_{1,3}+\theta^2_{2,3})d^5+o(d^5).\label{eq:detD4}
 \end{align}
 Substituting (\ref{eq:detD4}) into (\ref{eq:lowerbound2}) and invoking the asymptotic formula $\log(1-x)=-x+o(x)$ gives
 \begin{align*}
 r_{\{\{1,2\},\{3\}\}}(d)-r_C(d)\geq\frac{1}{2}(\theta^2_{1,3}+\theta^2_{2,3})d^2+o(d^2).
 \end{align*}
 This completes the proof of Lemma \ref{lem:3}.


\subsection{$L=3$ and $\mathbb{S}=\{\{1,2\},\{1,3\}\}$}

\begin{lemma}\label{lem:4}
For $d$ sufficiently close to 0,
\begin{align*}
&\hspace{0.05in}r_{\{\{1,2\},\{1,3\}\}}(d)=\min\limits_{d_1,d_2,d_3}r(d_1,d_2,d_3)\\
&\hspace{0.55in}\mbox{subject to}\quad d_{\ell}>0,\quad\ell=1,2,3,\nonumber\\
&\hspace{1.305in} d_2=d_3,\\
&\hspace{1.305in} d_1+d_2+d_3= 3d,\nonumber
\end{align*}
where
\begin{align*}
r(d_1,d_2,d_3)\triangleq\frac{1}{2}\log\Big(\frac{\det(\Gamma)}{2d_1d_2d_3}\Big(1+\sqrt{1+4\theta^2_{2,3}d_2d_3}\Big)\Big).
\end{align*}
\end{lemma}
\begin{IEEEproof}
	See Appendix \ref{app:2}.
	\end{IEEEproof}

Note that 
\begin{align}
&r(d_1,d_2,d_2)\nonumber\\
&=\frac{1}{2}\log\Big(\frac{\det(\Gamma)}{d_1d^2_2}\Big)+\frac{1}{2}\log\Big(1+\theta^2_{2,3}d^2_2+o(d^2_2)\Big)\nonumber\\
&=\frac{1}{2}\log\Big(\frac{\det(\Gamma)}{d_1d^2_2}\Big)+\frac{1}{2}\theta^2_{2,3}d^2_2+o(d^2_2).\label{eq:1213}
\end{align}
Consider the following convex optimization problem:
\begin{align*}
&\hspace{1.25in}\min\limits_{d_1,d_2}-\frac{1}{2}\log(d_1d^2_2)+\frac{1}{2}\theta^2_{2,3}d^2_2\\
&\hspace{0.55in}\mbox{subject to}\quad d_{\ell}>0,\quad\ell=1,2,\nonumber\\
&\hspace{1.305in} d_1+2d_2= 3d.\nonumber
\end{align*}
It can be readily shown that the optimizer satisfies
\begin{align*}
d_2=\begin{cases}
\frac{-1+\sqrt{1+4\theta^2_{2,3}d^2_1}}{2\theta^2_{2,3}d_1},&\theta_{2,3}\neq 0,\cr
d_1, &\theta_{2,3}=0.
\end{cases}
\end{align*}
In either case we have 
\begin{align*}
d_2=d_1-\theta^2_{2,3}d^3_1+o(d^3_1),
\end{align*}
which, together with the constraint $d_1+2d_2=3d$, implies
\begin{align}
&d_1=d+\frac{2}{3}\theta^2_{2,3}d^3+o(d^3),\label{eq:d1approx}\\
&d_2=d-\frac{1}{3}\theta^2_{2,3}d^3+o(d^3).\label{eq:d2approx}
\end{align}
Substituting (\ref{eq:d1approx}) and (\ref{eq:d2approx}) into  (\ref{eq:1213}) yields the desired result.

\subsection{$L=3$ and $\mathbb{S}=\{\{1,2\},\{1,3\},\{2,3\}\}$}

The desired result for this case is a simple consequence of the following lemma.
\begin{lemma}\label{lem:5}
For $d$ sufficiently close to 0,
\begin{align*}
r_{\{\{1,2\},\{1,3\},\{2,3\}\}}(d)=r_C(d).
\end{align*}
\end{lemma}
\begin{IEEEproof}
See Appendix \ref{app:3}.
\end{IEEEproof}

This completes the proof of Theorem \ref{thm:theorem1}.

\section{Conclusion}\label{sec:conclusion}

We have proposed a conjectural expression of the asymptotic gap between the rate-distortion function of an arbitrary generalized Gaussian multiterminal source coding system and that of its centralized counterpart in the high-resolution regime, and provided some supporting evidences by showing that this expression is valid when the number of sources is no more than 3. It is clear that the case-by-case study, as done in this work, is infeasible for proving the conjecture in its full generality, and a more conceptual approach is needed. We intend to give a more comprehensive treatment of this conjecture in a follow-up work by unifying and extending the existing achievability and converse arguments for multiterminal source coding using probabilistic graphical models. 




\appendices
\section{Proof of Lemma \ref{lem:1}}\label{app:1}

It can be deduced from  \cite[Theorem 6]{WCW10} that for $d$ sufficiently close to 0,
\begin{align}
&r_{\{\{1\},\{2\}\}}(d)=\min\limits_{d_1,d_2}r(d_1,d_2)\label{eq:min}\\
&\hspace{0.55in}\mbox{subject to}\quad d_{\ell}>0,\quad\ell=1,2,\nonumber\\
&\hspace{1.305in} d_1+d_2\leq 2d,\nonumber
\end{align}
where
\begin{align*}
r(d_1,d_2)\triangleq\frac{1}{2}\log\Big(\frac{\det(\Gamma)}{2d_1d_2}\Big(1+\sqrt{1+4\theta^2_{1,2}d_1d_2}\Big)\Big).
\end{align*}
One can readily prove Lemma \ref{lem:1} by observing that the minimum in (\ref{eq:min}) is achieved at $d_1=d_2=d$.

\section{Proof of Lemma \ref{lem:4}}\label{app:2}


The well-known Berger-Tung scheme \cite{Berger78, Tung78}  (see also \cite{XCWB07}) can be leveraged to establish the following upper bound on 
$r_{\mathbb{S}}(\cdot)$ for any cover $\mathbb{S}$ of $\{1,\cdots,L\}$.

\begin{proposition}\label{prop:BergerTung}
	For any Gaussian random variables/vectors $W_{\mathcal{S}}$, $\mathcal{S}\in\mathbb{S}$, jointly distributed with $(X_1,\cdots,X_L)$ such that $W_{\mathcal{S}}\leftrightarrow(X_{\ell})_{\ell\in\mathcal{S}}\leftrightarrow((X_{\ell'})_{\ell'\in\{1,\cdots,L\}\backslash\mathcal{S}}, (W_{\mathcal{S}'})_{\mathcal{S}'\in\mathbb{S}\backslash\mathcal{S}})$ form a Markov chain for any $\mathcal{S}\in\mathbb{S}$, we have
	\begin{align*}
	&r_{\mathbb{S}}\Big(\frac{1}{L}\mathrm{tr}(\mathrm{cov}((X_1,\cdots,X_L)|(W_{\mathcal{S}})_{\mathcal{S}\in\mathbb{S}}))\Big)\\
	&\leq \frac{1}{2}\log\Big(\frac{\det(\Gamma)}{\det(\mathrm{cov}((X_1,\cdots,X_L)|(W_{\mathcal{S}})_{\mathcal{S}\in\mathbb{S}}))}\Big).
	\end{align*}
\end{proposition}

Let $U_{\{1,2\}}$, $U_{\{1,3\}}$, $V_1$, $V_2$, and $V_3$ be defined as in Appendix \ref{app:3}.
The following facts can be verified via direct calculation.
\begin{itemize}
	\item[1)] The conditional joint distribution  of $U_{\{1,2\}}$, $U_{\{1,3\}}$, $V_1$, $V_2$, and $V_3$ given $(X_1,X_2,X_3)$ factors as
	\begin{align*}
	&p(u_{\{1,2\}},u_{\{1,3\}},v_1,v_2,v_3|x_1,x_2,x_3)\\
	&=p(u_{\{1,2\}}|x_1,x_2)p(u_{\{1,3\}}|x_1,x_3)\\
	&\quad\times p(v_1|x_1)p(v_2|x_2)p(v_3|x_3).
	\end{align*}
	
	\item[2)] The conditional joint distribution  of $X_1$, $X_2$, and $X_3$ given $(U_{\{1,2\}},U_{\{1,3\}},V_1,V_2,V_3)$ factors as 
	\begin{align*}
	&p(x_1,x_2,x_3|u_{\{1,2\}},u_{\{1,3\}},v_1,v_2,v_3)\\
	&=p(x_1|u_{\{1,2\}},u_{\{1,3\}},v_1)\\
	&\quad\times p(x_2,x_3|u_{\{1,2\}},u_{\{1,3\}},v_2,v_3).
	\end{align*}
\end{itemize}
Let $W_{\{1,2\}}\triangleq(U_{\{1,2\}},V_1,V_2)$  and $W_{\{1,3\}}\triangleq(U_{\{1,3\}},V_3)$. In light of the  above two facts, $W_{\{1,2\}}$ and $W_{\{1,3\}}$ satisfy the Markov chain constraints in Proposition \ref{prop:BergerTung} for $\mathbb{S}=\{\{1,2\},\{1,3\}\}$, and
\begin{align*}
&\mathrm{cov}((X_1,\cdots,X_L)|(W_{\mathcal{S}})_{\mathcal{S}\in\mathbb{S}})\\
&=\left(\begin{matrix}
\tilde{d}_1 & 0 & 0\\
0 & \tilde{d}_2 & \tilde{\rho}\sqrt{\tilde{d}_2\tilde{d}_3}\\
0 & \tilde{\rho}\sqrt{\tilde{d}_2\tilde{d}_3} & \tilde{d}_3
\end{matrix}\right),
\end{align*}
where
\begin{align*}
&\tilde{d}_1\triangleq\mathbb{E}[(X_1-\mathbb{E}[X_1|U_{\{1,2\}},U_{\{1,3\}},V_1])^2],\\
&\tilde{d}_2\triangleq\mathbb{E}[(X_2-\mathbb{E}[X_2|U_{\{1,2\}},U_{\{1,3\}},V_2,V_3])^2],\\
&\tilde{d}_3\triangleq\mathbb{E}[(X_3-\mathbb{E}[X_3|U_{\{1,2\}},U_{\{1,3\}},V_2,V_3])^2],\\
&\tilde{\rho}\triangleq\begin{cases}
\sqrt{1-\frac{2}{1+\sqrt{1+4\theta^2_{2,3}\tilde{d}_2\tilde{d}_3}}},& \theta_{2,3}\leq 0, \cr
-\sqrt{1-\frac{2}{1+\sqrt{1+4\theta^2_{2,3}\tilde{d}_2\tilde{d}_3}}},& \theta_{2,3}>0.
\end{cases}
\end{align*}
For any $d_{\ell}$, $\ell=1,2,3$, sufficiently close to 0, 
we can choose $\alpha_{\ell}$, $\ell=1,2,3$, such that $\tilde{d}_{\ell}=d_{\ell}$, $\ell=1,2,3$. Now invoking Proposition \ref{prop:BergerTung} shows that for $d$ sufficiently close to 0, 
\begin{align*}
&\hspace{0.05in}r_{\{\{1,2\},\{1,3\}\}}(d)\leq\min\limits_{d_1,d_2,d_3}r(d_1,d_2,d_3),
\end{align*}
where $d_{\ell}$, $\ell=1,2,3$, are subject to the constraints stated in Lemma \ref{lem:4}.

It remains to prove that this upper bound is in fact tight.
Consider two arbitrary encoding functions $\phi^{(n)}_{1}:\mathbb{R}^{2\times n}\rightarrow\mathcal{C}^{(n)}_1$ and $\phi^{(n)}_{2}:\mathbb{R}^{2\times n}\rightarrow\mathcal{C}^{(n)}_2$ satisfying 
\begin{align*}
\frac{1}{n}\sum\limits_{t=1}^n\mathbb{E}[(X_{\ell}(t)-\hat{X}_{\ell}(t))^2]\leq\hat{d}_{\ell},\quad\ell=1,2,3, 
\end{align*}
where 
\begin{align*}
&\hat{X}^n_{\ell}\triangleq\mathbb{E}[X^n_{\ell}|\phi^{(n)}_{1}(X^n_{1},X^n_2),\phi^{(n)}_2(X^n_1,X^n_3)],\\
&\hspace{1.9in} \ell=1,2,3.
\end{align*}
Note that
\begin{align}
&\sum\limits_{m=1}^2\log|\mathcal{C}^{(n)}_m|\nonumber\\
&\geq H(\phi^{(n)}_{1}(X^n_{1},X^n_2),\phi^{(n)}_2(X^n_1,X^n_3))\nonumber\\ &= I(X^n_1,X^n_2,X^n_3;\phi^{(n)}_{1}(X^n_{1},X^n_2),\phi^{(n)}_2(X^n_1,X^n_3))\nonumber\\
&=I(X^n_1;\phi^{(n)}_{1}(X^n_{1},X^n_2),\phi^{(n)}_2(X^n_1,X^n_3))\nonumber\\
&\quad+I(X^n_2,X^n_3;\phi^{(n)}_{1}(X^n_{1},X^n_2),\phi^{(n)}_2(X^n_1,X^n_3)|X^n_1).\label{eq:term12}
\end{align}
We have
\begin{align}
&I(X^n_1;\phi^{(n)}_{1}(X^n_{1},X^n_2),\phi^{(n)}_2(X^n_1,X^n_3))\nonumber\\
&\geq I(X^n_1;\hat{X}^n_1)\nonumber\\
&=\sum\limits_{t=1}^nI(X_1(t);\hat{X}^n_1|X^{t-1}_1)\nonumber\\
&=\sum\limits_{t=1}^nI(X_1(t);\hat{X}^n_1,X^{t-1}_1)\nonumber\\
&\geq\sum\limits_{t=1}^nI(X_1(t);\hat{X}_1(t))\nonumber\\
&\geq\sum\limits_{t=1}^n\frac{1}{2}\log\Big(\frac{\gamma_{1,1}}{\mathbb{E}[(X_1(t)-\hat{X}_1(t))^2]}\Big)\nonumber\\
&\geq\frac{n}{2}\log\Big(\frac{\gamma_{1,1}}{\frac{1}{n}\sum_{t=1}^n\mathbb{E}[(X_1(t)-\hat{X}_1(t))^2]}\Big)\nonumber\\
&\geq\frac{n}{2}\log\Big(\frac{\gamma_{1,1}}{\hat{d}_1}\Big).\label{eq:term1}
\end{align}
Now let
\begin{align*} &\phi^{(n)}_1(X^n_1,X^n_2)\triangleq\phi^{(n)}_1(Y^n_2|X^n_1),\\ &\phi^{(n)}_2(X^n_1,X^n_3)\triangleq\phi^{(n)}_2(Y^n_3|X^n_1),
\end{align*}
where
\begin{align*}
Y^n_{\ell}\triangleq X^n_{\ell}-\mathbb{E}[X^n_{\ell}|X^n_1],\quad \ell=2,3.
\end{align*}
Clearly, 
\begin{align}
&I(X^n_2,X^n_3;\phi^{(n)}_1(X^n_1,X^n_2),\phi^{(n)}_2(X^n_1,X^n_3)|X^n_1)\nonumber\\
&=I(Y^n_2,Y^n_3;\phi^{(n)}_1(Y^n_2|X^n_1),\phi^{(n)}_2(Y^n_3|X^n_1)|X^n_1).\label{eq:comb1}
\end{align}
Note that given $X^n_1$, $\phi^{(n)}_1(Y^n_2|X^n_1)\leftrightarrow Y^n_2\leftrightarrow Y^n_3\leftrightarrow\phi^{(n)}_2(Y^n_3|X^n_1)$ form a Markov chain. 
This observation suggests that one can establish a lower bound on $I(Y^n_2,Y^n_3;\phi^{(n)}_{1}(Y^n_2|x^n_1),\phi^{(n)}_2(Y^n_3|x^n_1)|X^n_1=x^n_1)$ by leveraging the converse arguments  developed for characterizing the minimum achievable sum-rate of quadratic Gaussian two-terminal source coding with source covariance matrix $\mathrm{cov}((X_2,X_3)|X_1)$ under  distortion constraints $\delta_2(x^n_1)$ and $\delta_3(x^n_1)$, where
\begin{align*}
&\delta_{\ell}(x^n_1)\triangleq\frac{1}{n}\sum\limits_{t=1}^n\mathbb{E}[(Y_{\ell}(t)-\tilde{Y}_{\ell}(t))^2|X^n_1=x^n_1],\\
&\hspace{2.5in}\ell=2,3,
\end{align*}
with
\begin{align*}
&\tilde{Y}^n_{\ell}\triangleq\mathbb{E}[Y^n_{\ell}|X^n_1,\phi^{(n)}_{1}(Y^n_2|X^n_1),\phi^{(n)}_2(Y^n_3|X^n_1)],\\
&\hspace{2.5in}\ell=2,3.
\end{align*}
Specifically, we have \cite{WTV08, WCW10, Oohama14, Courtade18, WC19}
\begin{align}
&I(Y^n_2,Y^n_3;\phi^{(n)}_1(Y^n_2|x^n_1),\phi^{(n)}_2(Y^n_3|x^n_1)|X^n_1=x^n_1)\nonumber\\
&\geq n\tilde{r}(\delta_2(x^n_1),\delta_3(x^n_1)), \label{eq:comb2}
\end{align}
where
\begin{align*}
&\tilde{r}(\delta_2,\delta_3)\\
&\triangleq\begin{cases}
&\frac{1}{2}\log\Big(\frac{\det(\Gamma)}{2\gamma_{1,1}\delta_2\delta_3}\Big(1+\sqrt{1+4\theta^2_{2,3}\delta_2\delta_3}\Big)\Big),\\
&\quad \max\Big\{\frac{\delta_2}{\gamma_{2,2|1}},\frac{\delta_3}{\gamma_{3,3|1}}\Big\}\leq\min\Big\{1,\frac{\gamma_{2,2|1}\gamma_{3,3|1}-\gamma^2_{2,3|1}}{\gamma_{2,2|1}\gamma_{3,3|1}}\\
&\hspace{0.97in}+\frac{\gamma^2_{2,3|1}}{\gamma_{2,2|1}\gamma_{3,3|1}}\min\Big\{\frac{\delta_2}{\gamma_{2,2|1}},\frac{\delta_3}{\gamma_{3,3|1}}\Big\}\Big\},\\
&\frac{1}{2}\log\Big(\min\Big\{1, \frac{\gamma_{2,2|1}}{\delta_2},\frac{\gamma_{3,3|1}}{\delta_3}\Big\}\Big),\quad\mbox{otherwise},
\end{cases}
\end{align*}
for $\delta_{\ell}>0$, $\ell=2,3$, with $\gamma_{2,2|1}$, $\gamma_{3,3|1}$, and $\gamma_{2,3|1}$ defined in (\ref{eq:gamma22}), (\ref{eq:gamma33}), and (\ref{eq:gamma23}), respectively.
Note that
\begin{align}
\mathbb{E}[\delta_{\ell}(X^n_1)]&=\frac{1}{n}\sum\limits_{t=1}^n\mathbb{E}[(Y_{\ell}(t)-\tilde{Y}_{\ell}(t))^2]\nonumber\\
&=\frac{1}{n}\sum\limits_{t=1}^n\mathbb{E}[(X_{\ell}(t)-\tilde{X}_{\ell}(t))^2]\nonumber\\
&\leq\frac{1}{n}\sum\limits_{t=1}^n\mathbb{E}[(X_{\ell}(t)-\hat{X}_{\ell}(t))^2]\nonumber\\
&\leq\hat{d}_{\ell},\quad\ell=2,3,\label{eq:relax}
\end{align}
where
\begin{align*}
&\tilde{X}^n_{\ell}\triangleq\mathbb{E}[X^n_{\ell}|X^n_1,\phi^{(n)}_{1}(X^n_{1},X^n_2),\phi^{(n)}_2(X^n_1,X^n_3)],\\
&\hspace{2.3in} \ell=2,3.
\end{align*}
Since $\tilde{r}(\delta_2,\delta_3)$ is a convex and monotonically decreasing function of $(\delta_2,\delta_3)$, it follows by (\ref{eq:relax}) that
\begin{align}
\mathbb{E}[\tilde{r}(\delta_2(X^n_1),\delta_3(X^n_1))]\geq \tilde{r}(\hat{d}_2,\hat{d}_3).\label{eq:comb3}
\end{align}
Combining (\ref{eq:comb1}), (\ref{eq:comb2}), and (\ref{eq:comb3}) shows that for $\hat{d}_2$ and $\hat{d}_3$ sufficiently close to 0,
\begin{align}
&I(X^n_2,X^n_3;\phi^{(n)}_{1}(X^n_{1},X^n_2),\phi^{(n)}_2(X^n_1,X^n_3)|X^n_1)\nonumber\\
&\geq\frac{n}{2}\log\Big(\frac{\det(\Gamma)}{2\gamma_{1,1}\hat{d}_2\hat{d}_3}\Big(1+\sqrt{1+4\theta^2_{2,3}\hat{d}_2\hat{d}_3}\Big)\Big).\label{eq:term2}
\end{align}
Substituting (\ref{eq:term1}) and (\ref{eq:term2}) into (\ref{eq:term12}) yields
\begin{align*}
\sum\limits_{m=1}^2\log|\mathcal{C}^{(n)}_m|\geq nr(\hat{d}_1,\hat{d}_2,\hat{d}_3)
\end{align*}
for $\hat{d}_{\ell}$, $\ell=1,2,3$, sufficiently close to 0,
which, together with a simple continuity argument, implies that for $d$ sufficiently close to 0,
\begin{align*}
&\hspace{0.05in}r_{\{\{1,2\},\{1,3\}\}}(d)\geq\min\limits_{d_1,d_2,d_3}r(d_1,d_2,d_3)\\
&\hspace{0.55in}\mbox{subject to}\quad d_{\ell}>0,\quad\ell=1,2,3,\nonumber\\
&\hspace{1.305in} d_1+d_2+d_3\leq 3d.\nonumber
\end{align*}
Clearly,  there is no loss of optimality in assuming that $d_2=d_3=d$ and $d_1+d_2+d_3=3d$. This completes the proof of Lemma \ref{lem:4}.



\section{Proof of Lemma \ref{lem:5}}\label{app:3}



Let $N_{\{1,2\}}$,  $N_{\{1,3\}}$, $N_{\{2,3\}}$, $Z_1$, $Z_2$, and $Z_3$ be zero-mean unit-variance Gaussian random variables. They are assumed to be mutually independent and independent of $(X_1,\cdots,X_L)$ as well.
For any $\lambda\in(0,1)$, let
\begin{align*}
&U_{\{1,2\}}\triangleq\begin{cases}
(1-\lambda)X_1+\lambda X_2+\eta_{1,2}N_{\{1,2\}}, &\theta_{1,2}<0,\cr 0, & \theta_{1,2}=0,\cr
(1-\lambda)X_1-\lambda X_2+\eta_{1,2}N_{\{1,2\}}, &\theta_{1,2}>0,
\end{cases}\\
&U_{\{1,3\}}\triangleq\begin{cases}
(1-\lambda)X_3+\lambda X_1+\eta_{1,3}N_{\{1,3\}}, &\theta_{1,3}<0,\cr 0, & \theta_{1,3}=0,\cr
(1-\lambda)X_3-\lambda X_1+\eta_{1,3}N_{\{1,3\}}, &\theta_{1,3}>0,
\end{cases}\\
&U_{\{2,3\}}\triangleq\begin{cases}
(1-\lambda)X_2+\lambda X_3+\eta_{2,3}N_{\{2,3\}}, &\theta_{2,3}<0,\cr 0, & \theta_{2,3}=0,\cr
(1-\lambda)X_2-\lambda X_3+\eta_{2,3}N_{\{1,2\}}, &\theta_{2,3}>0,
\end{cases}
\end{align*}
where
\begin{align*}
&\eta_{1,2}\triangleq\sqrt{\frac{(1-\lambda)\lambda(\gamma_{1,1|3}\gamma_{2,2|3}-\gamma^2_{1,2|3})}{|\gamma_{1,2|3}|}},\\
&\eta_{1,3}\triangleq\sqrt{\frac{(1-\lambda)\lambda(\gamma_{1,1|2}\gamma_{3,3|2}-\gamma^2_{1,3|2})}{|\gamma_{1,3|2}|}},\\
&\eta_{2,3}\triangleq\sqrt{\frac{(1-\lambda)\lambda(\gamma_{2,2|1}\gamma_{3,3|1}-\gamma^2_{2,3|1})}{|\gamma_{2,3|1}|}}
\end{align*}
with
\begin{align}
&\gamma_{1,1|3}\triangleq\frac{\theta_{2,2}}{\theta_{1,1}\theta_{2,2}-\theta^2_{1,2}},\nonumber\\
&\gamma_{2,2|3}\triangleq\frac{\theta_{1,1}}{\theta_{1,1}\theta_{2,2}-\theta^2_{1,2}},\nonumber\\
&\gamma_{1,2|3}\triangleq-\frac{\theta_{1,2}}{\theta_{1,1}\theta_{2,2}-\theta^2_{1,2}},\nonumber\\
&\gamma_{1,1|2}\triangleq\frac{\theta_{3,3}}{\theta_{1,1}\theta_{3,3}-\theta^2_{1,3}},\nonumber\\
&\gamma_{3,3|2}\triangleq\frac{\theta_{1,1}}{\theta_{1,1}\theta_{3,3}-\theta^2_{1,3}},\nonumber\\
&\gamma_{1,3|2}\triangleq-\frac{\theta_{1,3}}{\theta_{1,1}\theta_{3,3}-\theta^2_{1,3}},\nonumber\\
&\gamma_{2,2|1}\triangleq\frac{\theta_{3,3}}{\theta_{2,2}\theta_{3,3}-\theta^2_{2,3}},\label{eq:gamma22}\\
&\gamma_{3,3|1}\triangleq\frac{\theta_{2,2}}{\theta_{2,2}\theta_{3,3}-\theta^2_{2,3}},\label{eq:gamma33}\\
&\gamma_{2,3|1}\triangleq-\frac{\theta_{2,3}}{\theta_{2,2}\theta_{3,3}-\theta^2_{2,3}}.\label{eq:gamma23}
\end{align}
Moreover, for any $\alpha_{\ell}\geq 0$, $\ell=1,2,3$, let
\begin{align*}
&V_1\triangleq \alpha_1X_1+Z_1,\\
&V_2\triangleq\alpha_2X_2+Z_2,\\
&V_3\triangleq\alpha_3X_3+Z_3.
\end{align*}
The following facts can be verified via direct calculation.
\begin{itemize}
	\item[1)] The conditional joint distribution  of $U_{\{1,2\}}$, $U_{\{1,3\}}$, $U_{\{2,3\}}$, $V_1$, $V_2$, and $V_3$ given $(X_1,X_2,X_3)$ factors as
\begin{align*}
&p(u_{\{1,2\}},u_{\{1,3\}},u_{\{2,3\}},v_1,v_2,v_3|x_1,x_2,x_3)\\
&=p(u_{\{1,2\}}|x_1,x_2)p(u_{\{1,3\}}|x_1,x_3)p(u_{\{2,3\}}|x_2,x_3)\\
&\quad\times p(v_1|x_1)p(v_2|x_2)p(v_3|x_3).
\end{align*}

\item[2)] The conditional joint distribution  of $X_1$, $X_2$, and $X_3$ given $(U_{\{1,2\}},U_{\{1,3\}},U_{\{2,3\}},V_1,V_2,V_3)$ factors as 
\begin{align*}
&p(x_1,x_2,x_3|u_{\{1,2\}},u_{\{1,3\}},u_{\{2,3\}},v_1,v_2,v_3)\\
&=p(x_1|u_{\{1,2\}},u_{\{1,3\}},v_1)p(x_2|u_{\{1,2\}},u_{\{2,3\}},v_2)\\
&\quad\times p(x_3|u_{\{1,3\}},u_{\{2,3\}},v_3).
\end{align*}
\end{itemize}
Let $W_{\{1,2\}}\triangleq(U_{\{1,2\}},V_1)$,  $W_{\{1,3\}}\triangleq(U_{\{1,3\}},V_3)$, and $W_{\{2,3\}}\triangleq(U_{\{2,3\}},V_2)$. In light of the  above two facts, $W_{\{1,2\}}$, $W_{\{1,3\}}$, and $W_{\{2,3\}}$ satisfy the Markov chain constraints in Proposition \ref{prop:BergerTung} for $\mathbb{S}=\{\{1,2\},\{1,3\},\{2,3\}\}$, and
\begin{align*}
\mathrm{cov}((X_1,\cdots,X_L)|(W_{\mathcal{S}})_{\mathcal{S}\in\mathbb{S}})=\left(\begin{matrix}
\bar{d}_1 & 0 & 0\\
0 & \bar{d}_2 & 0\\
0 & 0 & \bar{d}_3
\end{matrix}\right),
\end{align*}
where 
\begin{align*}
&\bar{d}_1\triangleq\mathbb{E}[(X_1-\mathbb{E}[X_1|U_{\{1,2\}},U_{\{1,3\}},V_1])^2],\\
&\bar{d}_2\triangleq\mathbb{E}[(X_2-\mathbb{E}[X_2|U_{\{1,2\}},U_{\{2,3\}},V_2])^2],\\
&\bar{d}_3\triangleq\mathbb{E}[(X_3-\mathbb{E}[X_3|U_{\{1,3\}},U_{\{2,3\}},V_3])^2].
\end{align*}
For any $d$ sufficiently close to 0, 
we can choose $\alpha_{\ell}$, $\ell=1,2,3$, such that $\bar{d}_{\ell}=d$, $\ell=1,2,3$. Now invoking Proposition \ref{prop:BergerTung} completes the proof of Lemma \ref{lem:5}.

\end{document}